\documentclass[a4paper]{svmult}
\usepackage{times}
\usepackage[dvips]{graphicx,epsfig}
\usepackage{amsmath,amsfonts,amssymb}
\usepackage{cite,url}

\begin{document}

\frontmatter

%\tableofcontents

\mainmatter

\title*{Effective Interactions In Neutron-Rich Matter}

\author{\underline{{P.~G.~Krastev}}\inst{1} \and {F.~Sammarruca}\inst{2} \and {Bao-An
Li}\inst{1} \and {A.~Worley}\inst{1}}

\titlerunning{Effective Interactions In Neutron-Rich Matter}
\authorrunning{{P.~G.~Krastev}, {F.~Sammarruca}, {Bao-An Li}, and
{A.~ Worley}}

\toctitle{Effective Interactions In Neutron-Rich Matter}
\tocauthor{{P.~G.~Krastev}, {F.~Sammarruca}, {Bao-An Li}, and
{A.~Worley}}

\institute{{Texas A\&M University-Commerce, Commerce, TX 75429,
U.S.A.} \and {University of Idaho, Moscow, ID 83843, U.S.A.}}

\maketitle

\begin{abstract}
Properties of effective interactions in neutron-rich matter are
reflected in the medium's equation of state (EOS), which is a
relationship among several state variables. Spin and isospin
asymmetries play an important role in the energy balance and could
alter the stability conditions of the nuclear EOS. The EOS has
far-reaching consequences for numerous nuclear processes in both the
terrestrial laboratories and the cosmos. Presently the EOS,
especially for neutron-rich matter, is still very uncertain.
Heavy-ion reactions provide a unique means to constrain the EOS,
particularly the density dependence of the nuclear symmetry energy.
On the other hand, microscopic, self-consistent, and parameter-free
approaches are ultimately needed for understanding nuclear
properties in terms of the fundamental interactions among the basic
constituents of nuclear systems. In this talk, after a brief review
of our recent studies on spin-polarized neutron matter, we discuss
constraining the changing rate of the gravitational constant $G$ and
properties of (rapidly) rotating neutron stars by using a nuclear
EOS partially constrained by the latest terrestrial nuclear
laboratory data.
\end{abstract}

\section{Introduction}
\quad Properties of matter under extreme pressure and density are of
great interest in modern physics as they are closely related to
numerous important nuclear phenomena in both the terrestrial
laboratories and space. These properties depend on the interactions
among the constituents of matter and are reflected in the equation
of state (EOS) characterizing the medium. At high densities
non-nucleonic degrees of freedom appear gradually due to the rapid
rise of nucleon chemical potentials~\cite{BBS2000}. Among these
particles are strange hyperons such as $\Lambda^0$ and $\Sigma^-$.
At even higher densities matter is expected to undergo a phase
transition to quark-gluon plasma~\cite{Weber:1999a}. Extracting the
transition density from QCD lattice calculations is a formidable
problem which is still presently unsolved. These complications
introduce great challenges on our way to understanding behavior of
matter in terms of interactions among its basic ingredients.

The EOS is important for many key processes in both nuclear physics
and astrophysics. It has far-reaching consequences and governs
dynamics of supernova explosions, formation of heavy elements,
properties and structure of neutron stars, and the time variations
of the gravitational constant $G$. Presently, the detailed knowledge
of the EOS is still far from complete mainly due to the very poorly
known density dependence of the nuclear symmetry energy,
$E_{sym}(\rho)$. Different many-body theories yield, often, rather
controversial predictions for the trend of $E_{sym}(\rho)$ and thus
the EOS. On the other hand, heavy-ion reactions at intermediate
energies have already constrained significantly the density
dependence of $E_{sym}$ around nuclear saturation density, see
e.g.~\cite{Shi:2003np,Tsang,Chen:2004si,Steiner:2005rd,Li:2005jy,Chen:2005ti}.
Consequently, these constraints place also significant limits on the
possible configurations of both static~\cite{Li:2005sr} and
(rapidly) rotating~\cite{Krastev:2007wh} neutron stars, and the
possible time variations of $G$~\cite{Krastev:2007en}. In this
report we review these findings. We also revisit our recent studies
on spin-asymmetric neutron matter~\cite{Krastev:2006xn} with a
particular emphasis on high densities.

\section{Spin-polarized neutron matter: high-density regime}

\begin{figure}[b!]
\begin{center}
\includegraphics[width=6.cm,height=6.cm]{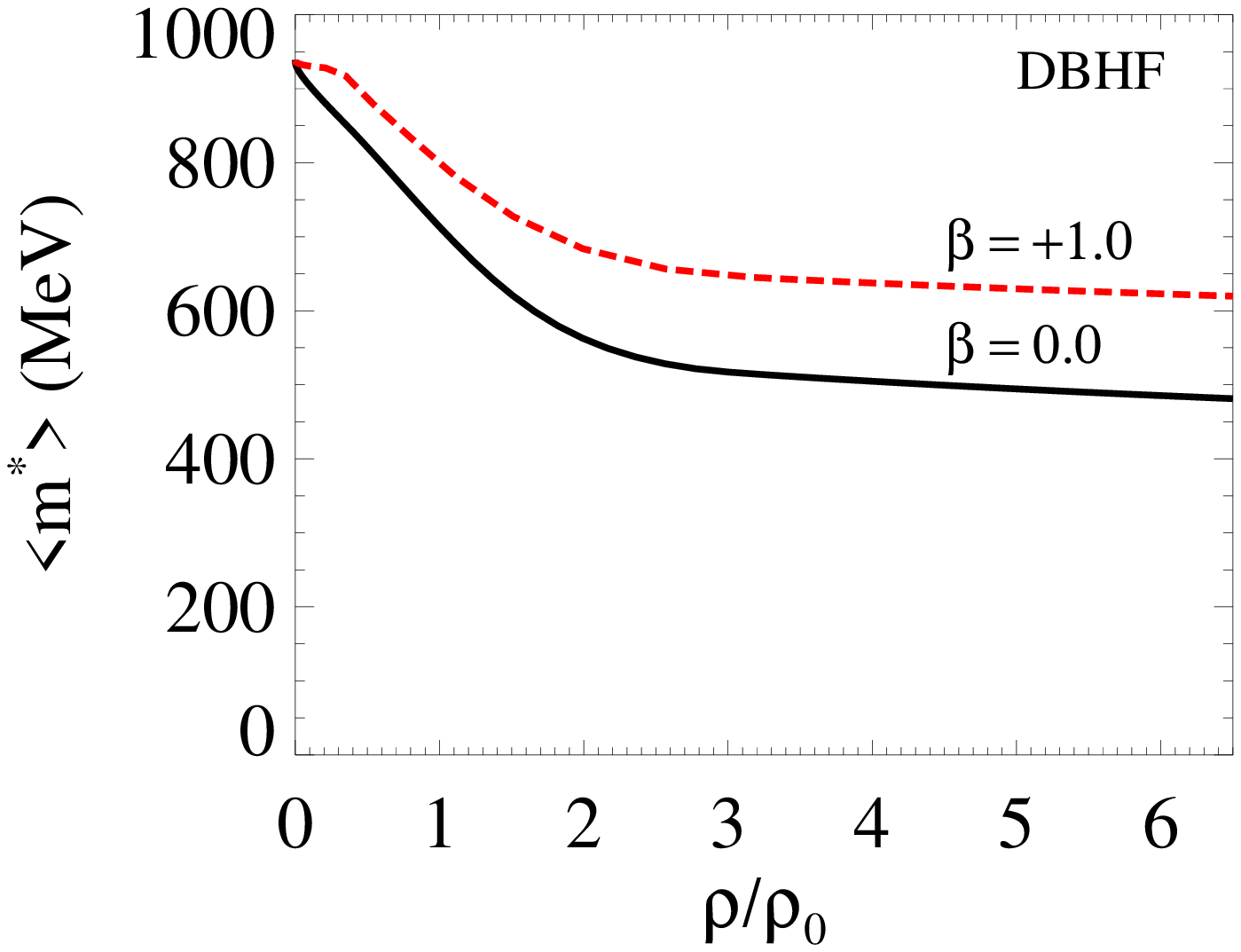}
\includegraphics[width=5.cm,height=6.cm]{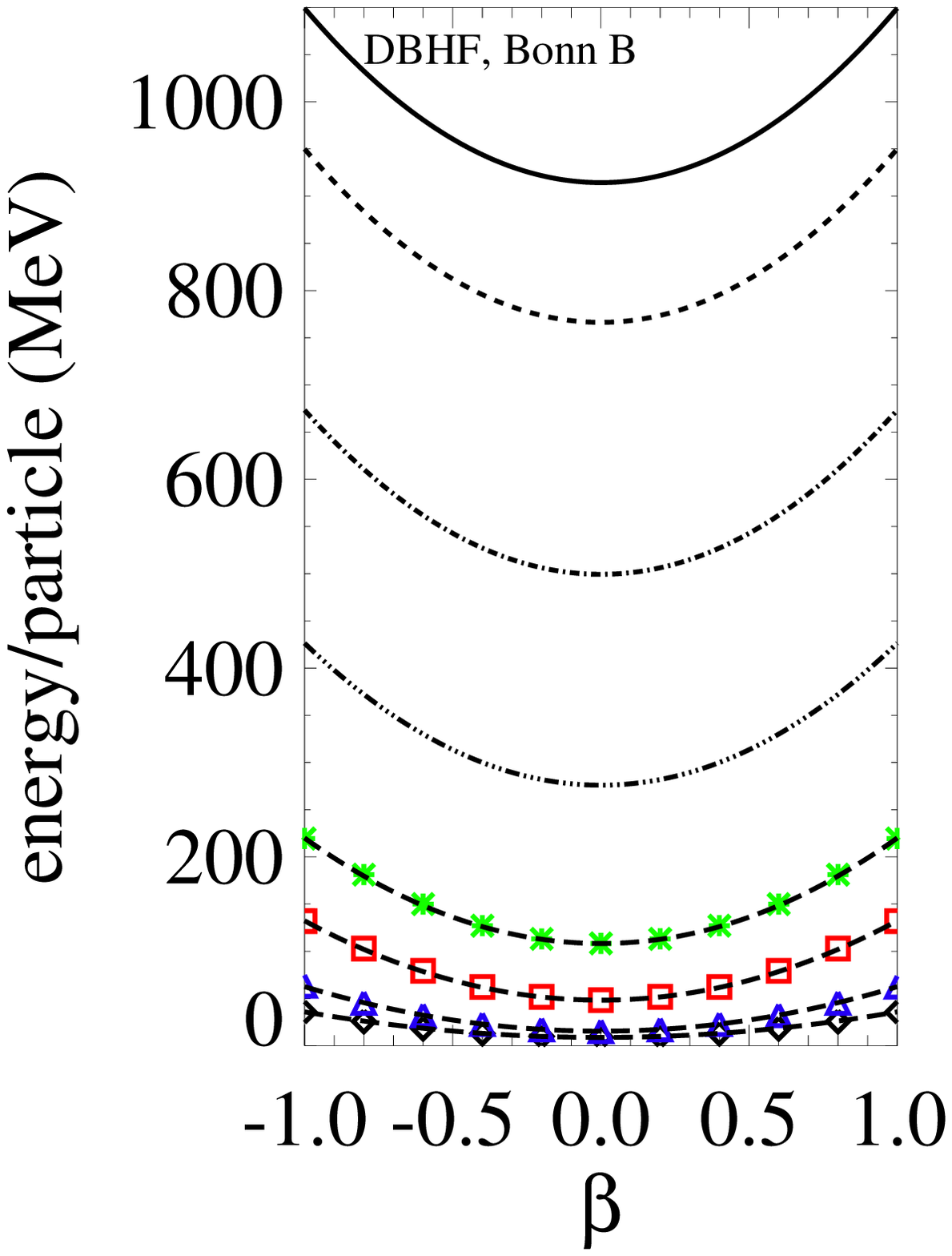}
\caption{{\protect\small Left panel: Neutron effective masses used
in the DBHF calculations of the EOS. The angular dependence is
averaged out; Right panel: Average energy per particle at densities
equal to 0.5, 1, 2, 3, 5, 7, 9, and 10 times $\rho_0$ (from lowest
to highest curve).}}
\end{center}
\label{mr}
\end{figure}

\quad Studies of magnetic properties of dense matter are of great
current interest in conjunction with studies of pulsars, which are
believed to be rotating neutron stars with strong surface magnetic
fields. Here we summarize briefly the results of our recent study on
properties of spin-polarized pure neutron matter. For a detailed
description of the calculation we refer the interested reader to
Ref.~\cite{Krastev:2006xn} and the references therein. The
computation is microscopic and treats the nucleons in the medium
relativistically. The starting point of every microscopic
calculation of nuclear structure and reactions is a realistic
nucleon-nucleon (NN) free-space interaction. A realistic and
quantitative model for the nuclear force with reasonable theoretical
foundations is the one-boson-exchange (OBE)
model~\cite{Machleidt89}. Our standard framework consists of the
Bonn B OBE potential together with the Dirac-Brueckner-Hartree-Fock
(DBHF) approach to nuclear matter. A detailed description of
applications of the DBHF method to isospin symmetric and asymmetric
matter, and neutron-star properties can be found in
Refs.~\cite{Alonso:2003aq,Sammarruca:2004cy,Sammarruca:2006jd,Krastev:2006ii}.

\begin{figure}[t!]
\begin{center}
\includegraphics[width=5.cm,height=5.cm]{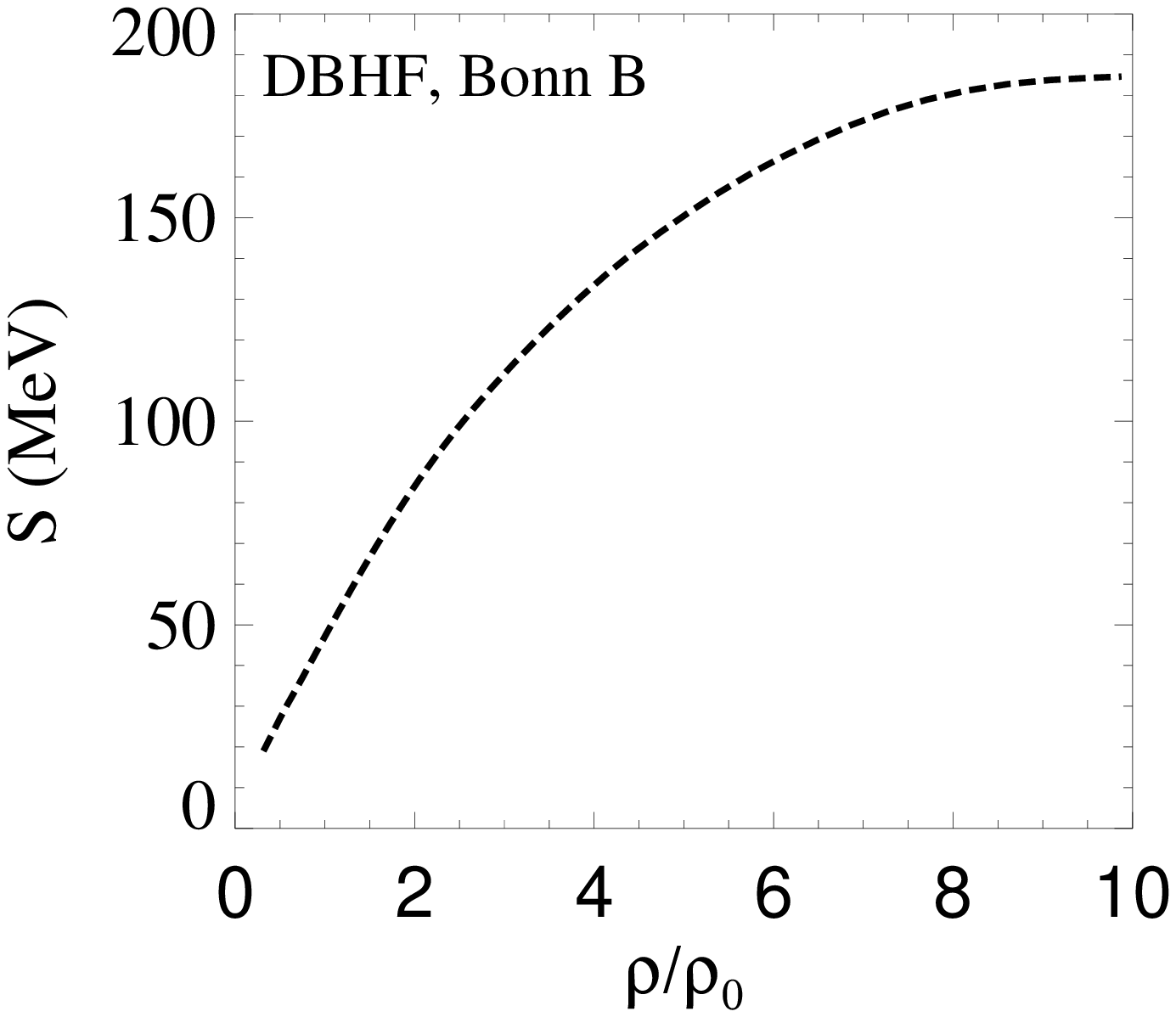}
\includegraphics[width=5.cm,height=5.cm]{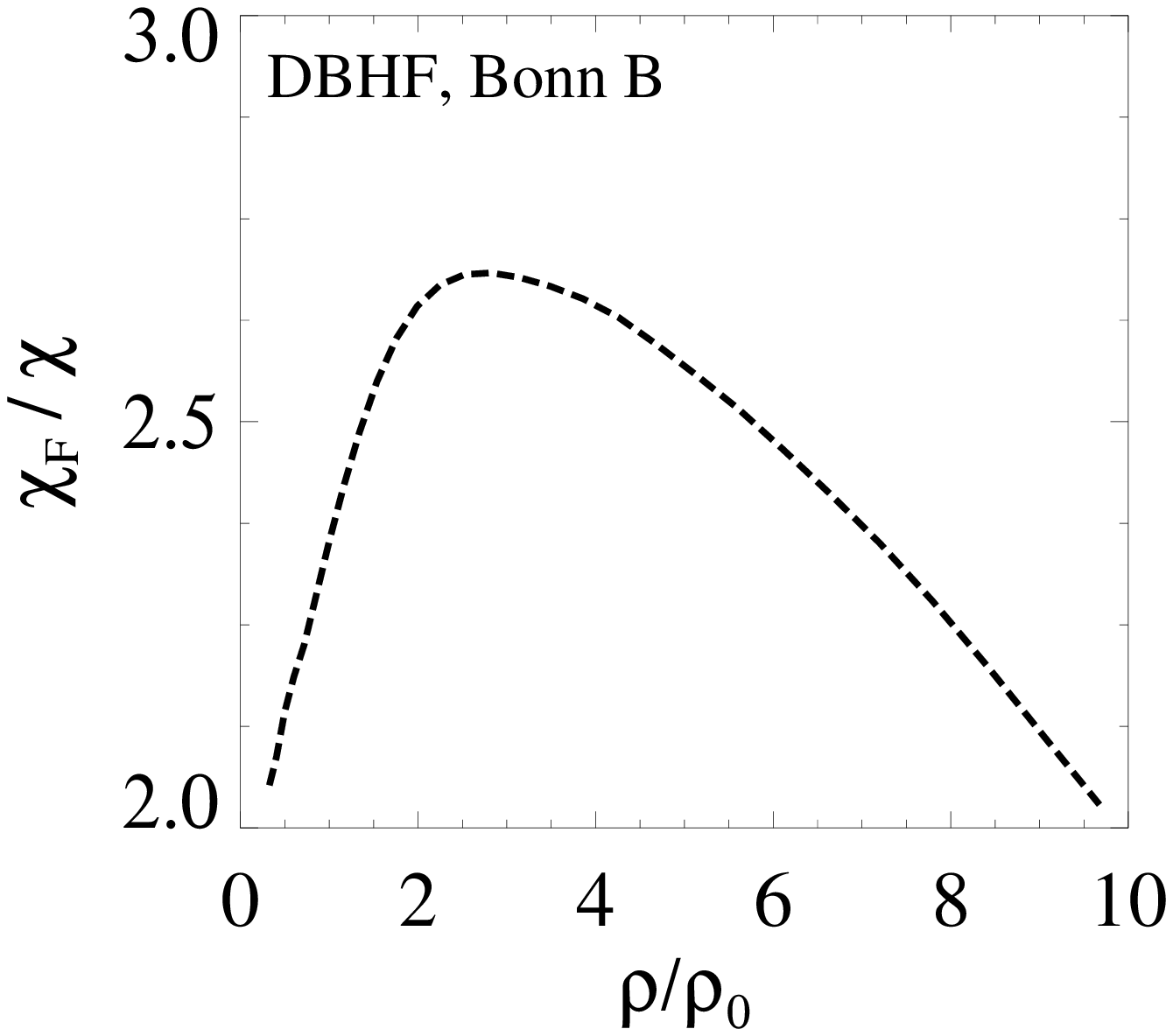}
\caption{{\protect\small Left panel: Density dependence of the spin
symmetry energy obtained with the DBHF model. Right panel: Density
dependence of the ratio $\chi_F /\chi $.}}
\end{center}
\label{mr}
\end{figure}

To explore the possibility for a ferromagnetic transition the DBHF
calculation has been extended to densities as high as $10\rho_0$
(with $\rho_0\approx 0.16fm^{-3}$ the density of normal nuclear
matter). Here we recall that the onset of ferromagnetic
instabilities above a given density would imply that the
energy-per-particle of a completely polarized state is lower than
the one of unpolarized neutron matter. The same method as the one
used in Ref.~\cite{Krastev:2006ii} has been applied to obtain the
energy-per-particle where a self-consistent solution cannot be
obtained (see Section III of Ref.~\cite{Krastev:2006ii} for
details). The (angle-averaged) neutron effective masses for both the
unpolarized and fully polarized case are shown in Fig.~1 (left
panel) as a function of density. The spin-asymmetry parameter,
$\beta=(\rho^{\uparrow}-\rho^{\downarrow})/(\rho^{\uparrow}+\rho^{\downarrow})$,
quantifies the degree of asymmetry of the system. It can take values
between -1 and +1 with 0 and the limits $\pm 1$ corresponding to
unpolarized and completely polarized matter respectively.
$\rho^{\uparrow}$ and $\rho^{\downarrow}$ are the densities of
neutrons with spins up/down. DBHF predictions for the average energy
per particle are shown in Fig.~1 (right panel) at densities ranging
from $\rho=0.5\rho_0$ to $10\rho_0$. What we observe is best seen
through the density dependence of spin-symmetry energy, $S(\rho)$,
which is the difference between the energies of completely polarized
and unpolarized neutron matter
\begin{equation}\label{eq:1}
S(\rho)=\bar{e}(\rho,\beta=1)-\bar{e}(\rho,\beta=0)
\end{equation}
A negative sign of $S(\rho)$ would signify that a polarized system
is more stable than unpolarized one. The spin-symmetry energy is
shown as a function of density in Fig~2 (left panel). We see that at
high density the energy shift between polarized and unpolarized
matter continues to grow, but at a smaller rate, and eventually
appear to saturate. For a detailed analysis of the observed behavior
of $S(\rho)$ see Ref.~\cite{Krastev:2006xn}. Here we should mention
that although the curvature of the spin-symmetry energy may suggest
that ferromagnetic instabilities are in principle possible within
the Dirac model, inspection of Fig.~2 reveals that such transition
does not take place at least up to 10$\rho_0$. Clearly, it would not
be appropriate to explore even higher densities without additional
considerations, such as transition to a quark phase. In fact, even
on the high side of the densities considered here, softening of the
equation of state from additional degrees of freedom not included in
the present model may be necessary in order to draw a more definite
conclusion. In the right panel of Fig.~2 we show the density
dependence of magnetic susceptibility, $\chi$, in terms of $\chi_F$,
the magnetic susceptibility of a free Fermi gas. $\chi(\rho)$ is
directly related to $S(\rho)$ through
\begin{equation}
\chi =\frac{ \mu^2 \rho}{2S(\rho)},
\end{equation}
with $\mu$ the neutron magnetic moment. Clearly, similar
observations apply to both left and right frames of Fig~2. (The
magnetic susceptibility would show an infinite discontinuity,
corresponding to a sign change of $S(\rho)$, in case of a
ferromagnetic instability.)

In summary of this section, the EOSs we obtain with the DBHF model
are generally rather repulsive at the larger densities. The energy
of the unpolarized system (where all $nn$ partial waves are
allowed), grows rapidly at high density with the result that the
energy difference between totally polarized and unpolarized neutron
matter tends to slow down with density. This may be interpreted as a
{\it precursor} of spin-separation instabilities, although no such
transition is actually seen up to 10$\rho_0$.

\section{Constraining a possible time variation of the gravitational constant $G$
with nuclear data from terrestrial laboratories}

\quad Testing the constancy of the gravitational constant $G$ is a
longstanding fundamental question in natural science. As first
suggested by Jofr\'{e}, Reisenegger and
Fern\'{a}ndez~\cite{Jofre:2006ug}, Dirac's
hypothesis~\cite{Dirac:1937ti} of a decreasing gravitational
constant $G$ with time due to the expansion of the Universe would
induce changes in the composition of neutron stars, causing
dissipation and internal heating. Eventually, neutron stars reach
their quasi-stationary states where cooling, due to neutrino and
photon emissions, balances the internal heating. The correlation of
surface temperatures and radii of some old neutron stars may thus
carry useful information about the rate of change of $G$. Using the
density dependence of the nuclear symmetry energy, constrained by
recent terrestrial laboratory data on isospin diffusion in heavy-ion
reactions at intermediate
energies~\cite{Shi:2003np,Tsang,Chen:2004si,Steiner:2005rd,Li:2005jy,Chen:2005ti},
and the size of neutron skin in
$^{208}Pb$~\cite{Steiner:2004fi,Horowitz:2000xj,Horowitz:2002mb,RP2005},
within the {\it gravitochemical heating} formalism developed by
Jofr\'{e} et al.~\cite{Jofre:2006ug}, we obtain an upper limit for
the relative time variation $|\dot{G}/G|$ in the range
$(4.5-21)\times 10^{-12}yr^{-1}$. In what follows we briefly review
our calculation. For details see Ref.~\cite{Krastev:2007en}.

Recently a new method, called {\it gravitochemical
heating}~\cite{Jofre:2006ug}, has been introduced to constrain a
hypothetical time variation in $G$, most frequently expressed as
$|\dot{G}/G|$. In Ref.~\cite{Jofre:2006ug} the authors suggested
that such a variation of the gravitational constant would perturb
the internal composition of a neutron star, producing entropy which
is partially released through neutrino emission, while a similar
fraction is eventually radiated as thermal photons. A constraint on
the time variation of $G$ is achieved via a comparison of the
predicted surface temperature with the available empirical value of
an old neutron star~\cite{Kargaltsev:2003eb}. The gravitochemical
heating formalism is based on the results of Fern\'{a}ndez and
Reisenegger~\cite{Fernandez:2005cg} (see
also~\cite{Reisenegger:1994be}) who demonstrated that internal
heating could result from spin-down compression in a rotating
neutron star ({\it rotochemical heating}). In both cases (gravito-
and rotochemical heatings) predictions rely heavily on the equation
of state (EOS) of stellar matter used to calculate the neutron star
structure. Accordingly, detailed knowledge of the EOS is critical
for setting a reliable constraint on the time variation of $G$.

\begin{figure}[t!]
\begin{center}
\includegraphics[width=5.cm,height=6.cm]{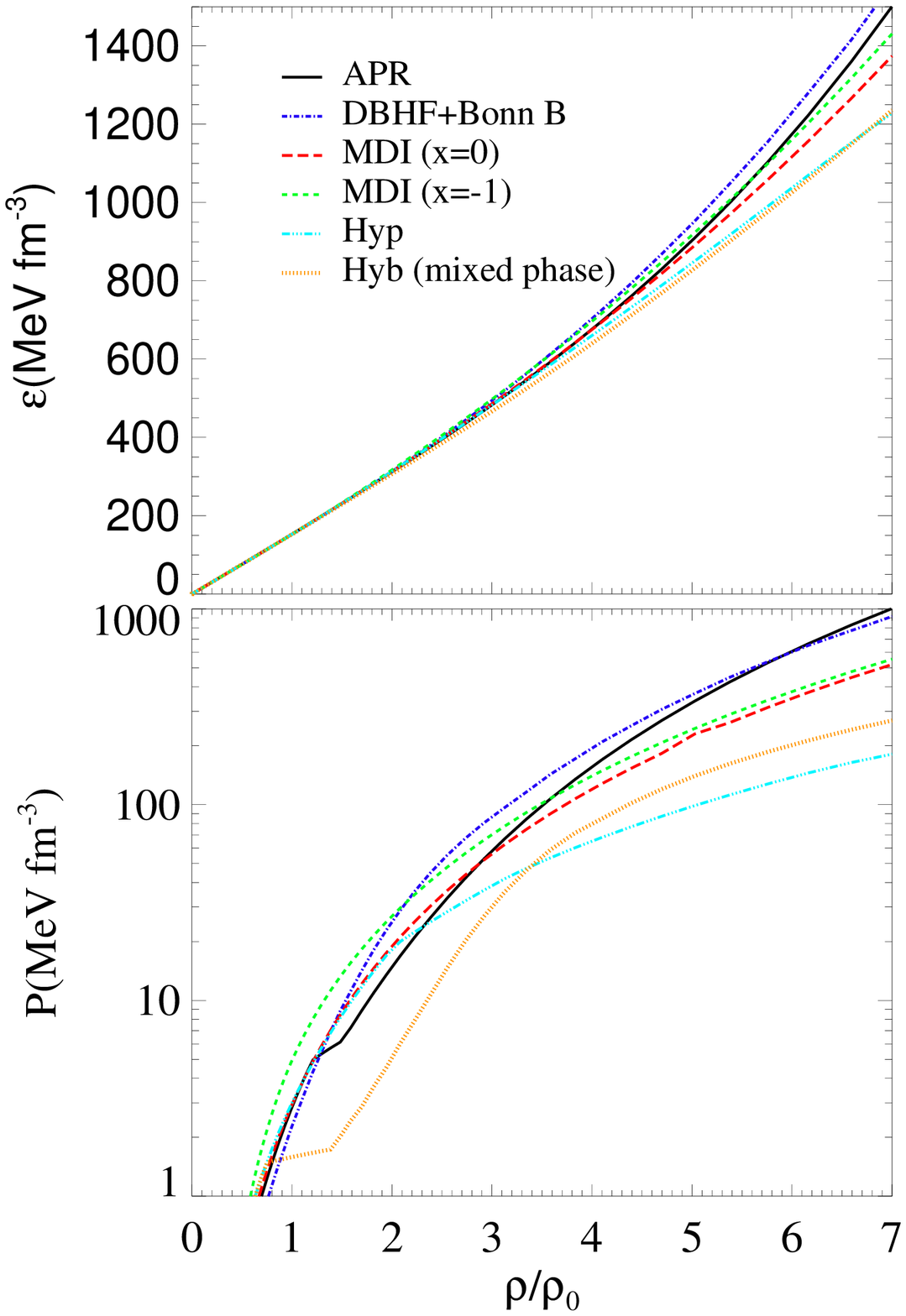}
\includegraphics[width=5.cm,height=6.cm]{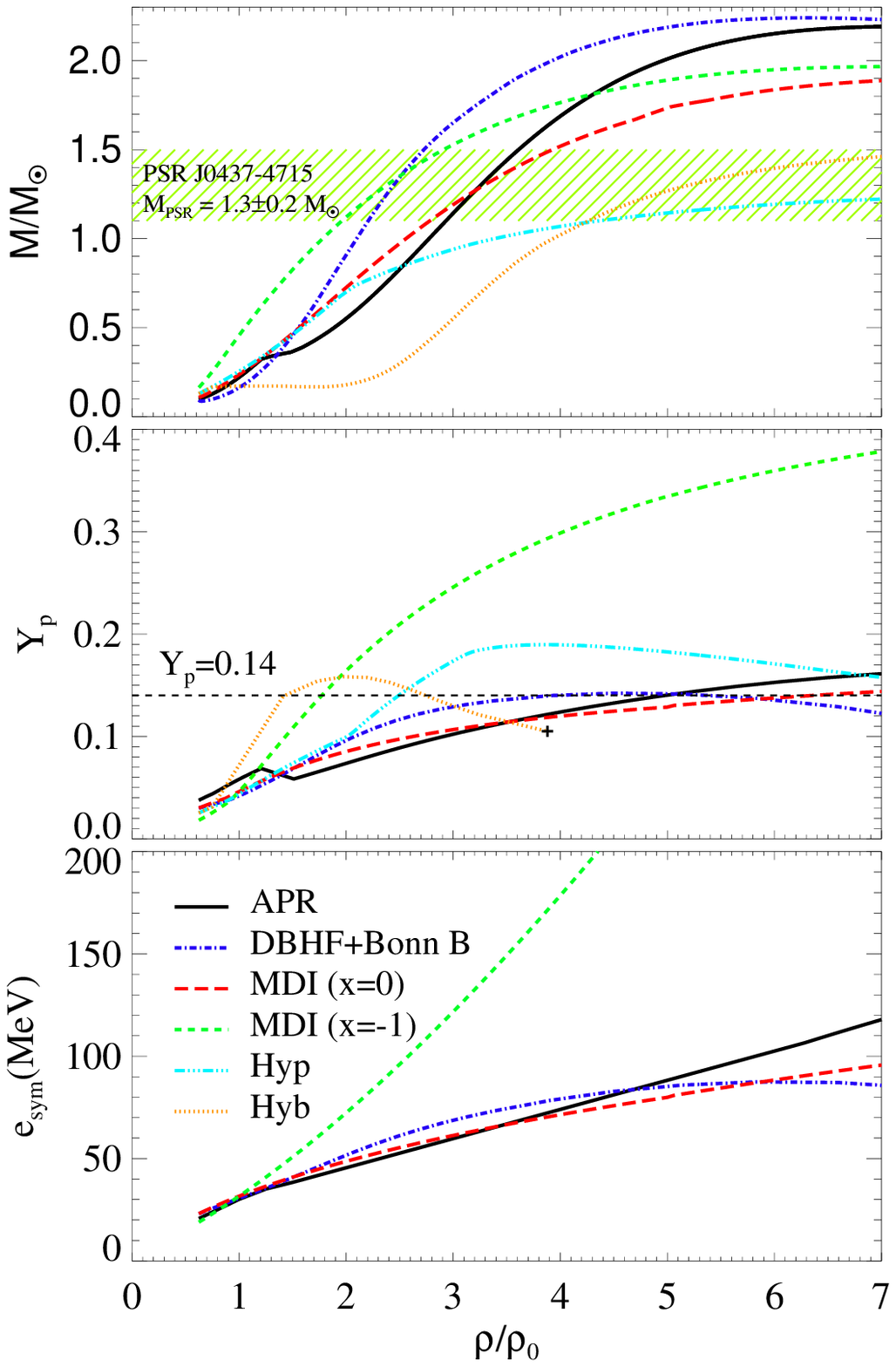}\vspace{5mm}
\caption{{\protect\small Left panel: Equation of state of stellar
matter in $\beta$-equilibrium. The upper panel shows the total
energy density and lower panel the pressure as function of the
baryon number density (in units of $\rho_0$); Right panel: Neutron
star mass, proton fraction, $Y_p$, and symmetry energy, $e_{sym}$.
The upper frame displays the neutron star mass as a function of
baryon number density. The middle frame shows the proton fraction
and the lower frame the nuclear symmetry energy as a function of
density. (Symmetry energy is shown for the nucleonic EOSs only.) The
proton fraction curve of the Hyb EOS is terminated at the beginning
of the quark phase. The termination point is denoted by a ``cross''
character.}}
\end{center}
\label{mr}
\end{figure}

Currently, theoretical predictions of the EOS of neutron-rich matter
diverge widely mainly due to the uncertain density dependence of the
nuclear symmetry energy. Consequently, to provide a stringent
constraint on the time variation of $G$, one should attempt to
reduce the uncertainty due to the $E_{sym}(\rho)$. Recently
available nuclear reaction data allowed us to constrain
significantly the density dependence of the symmetry energy mostly
in the sub-saturation density region. While high energy radioactive
beam facilities under construction will provide a great opportunity
to pin down the high density behavior of the nuclear symmetry energy
in the future. We apply the gravitochemical method with several EOSs
describing matter of purely nucleonic ($npe\mu$) as wells as
hyperonic and hybrid stars. Among the nucleonic matter EOSs, we pay
special attention to the one calculated with the MDI
interaction~\cite{Das:2002fr}. The symmetry energy $E_{sym}(\rho)$
of the MDI EOS is constrained in the sub-saturation density region
by the available nuclear laboratory data, while in the high-density
region we assume a continuous density functional. The EOS of
symmetric matter for the MDI interaction is constrained up to about
five times the normal nuclear matter density by the available data
on collective flow in relativistic heavy-ion reactions.

\begin{figure}[t!]
\begin{center}
\includegraphics[width=5.8cm,height=4.5cm]{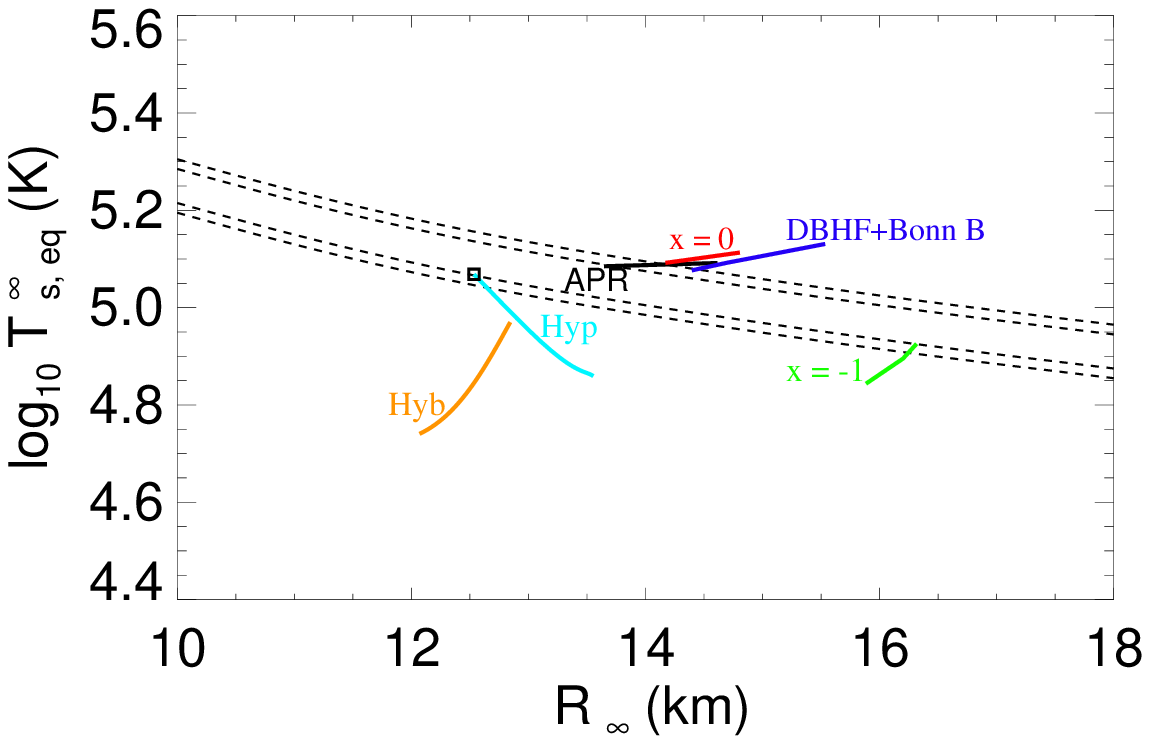}
\includegraphics[width=5.8cm,height=4.5cm]{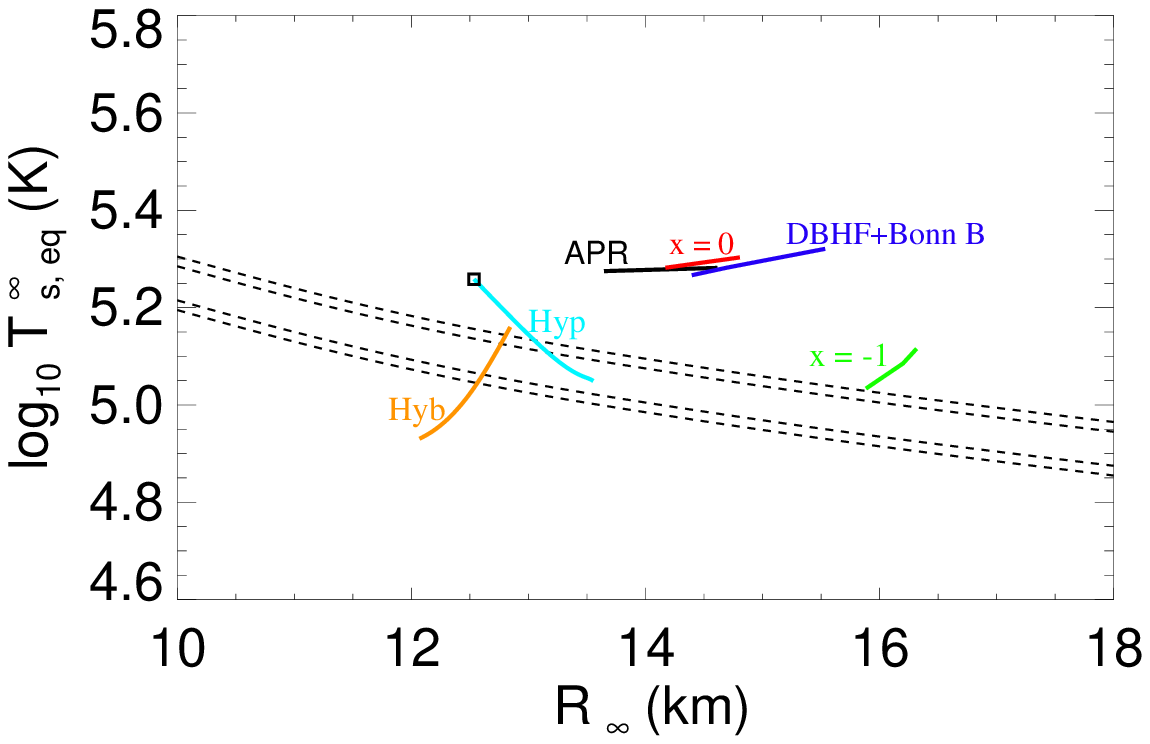}
\caption{{\protect\small Left panel: Neutron star stationary surface
temperature for stellar models satisfying the mass constraint by
Hotan et al.~\cite{HBO2006}. The solid lines are the predictions
versus the stellar radius for the considered neutron star sequences.
Dashed lines correspond to the 68\% and 90\% confidence contours of
the black-body fit of Kargaltsev et al.~\cite{Kargaltsev:2003eb}.
The value of $|\dot{G}/G|=4.5\times 10^{-12}yr^{-1}$ is chosen so
that predictions from the $x=0$ EOS are just above the observational
constraints; Right panel: Same as left panel but assuming
$|\dot{G}/G|=2.1\times 10^{-11}yr^{-1}$.}}
\end{center}
\label{mr}
\end{figure}

The EOSs applied with the gravitochemical heating method are shown
in Fig.~3 (left panel). For description of these EOSs see
Ref.~\cite{Krastev:2007en} and references therein. The parameter $x$
is introduced in the MDI interaction to reflect the largely
uncertain density dependence of the $E_{sym}(\rho)$ as predicted by
various many-body approaches. Since, as demonstrated in
Refs.~\cite{Li:2005jy,Li:2005sr}, only equations of state with $x$
between -1 and 0 have symmetry energies consistent with the isospin
diffusion data and measurements of the skin thickness of $^{208}Rb$,
we thus consider only these two limiting cases. Fig.~3 (right panel)
displays the neutron star mass (upper frame), the proton fraction
(middle frame) and the nuclear symmetry energy (lower frame). The
shaded region in the upper frame corresponds to the mass constraint
by Hotan et al.~\cite{HBO2006}.

As shown in Ref.~\cite{Jofre:2006ug} the stationary surface
temperature is directly related to the relative changing rate of $G$
via $ T_s^{\infty}=\tilde{\cal
D}\left|\frac{\dot{G}}{G}\right|^{2/7}$, where the function
$\tilde{\cal D}$ is a quantity depending only on the stellar model
and the equation of state. The correlation of surface temperatures
and radii of some old neutron stars may thus carry useful
information about the changing rate of $G$. Using the constrained
symmetry energy with $x=0$ and $x=-1$ shown in Fig.~3 (right panel),
within the gravitochemical heating formalism, as shown in Fig.~4, we
obtained an upper limit of the relative changing rate of $G$ in the
range of $(4.5-21)\times 10^{-12}yr^{-1}$. This is the best
available estimate in the literature~\cite{Krastev:2007en}. For a
comparison, results with the EOS from recent DBHF
calculations~\cite{Alonso:2003aq,Krastev:2006ii} with the Bonn B OBE
potential are also shown. Predictions with the DBHF+Bonn B EOS give
roughly the same value for the stationary surface temperature, but
at slightly larger neutron-star radius relative to the $x=0$ EOS.
For the effect of hyperonic and quark phases of matter on the
possible time variations of $G$ we refer the reader to our analysis
in Ref.~\cite{Krastev:2007en}.

The gravitochemical heating mechanism has the potential to become a
powerful tool for constraining gravitational physics. Since the
method relies on the detailed neutron star structure, which, in
turn, is determined by the EOS of stellar matter, further progress
in our understanding of properties of dense, neutron-rich matter
will make this approach more effective.

\section{Constraining properties and structure of rapidly rotating neutron stars}

\quad Because of their strong gravitational binding neutron stars
can rotate very fast~\cite{Bejger:2006hn}. The first millisecond
pulsar PSR1937+214, spinning at $\nu=641 Hz$~\cite{Backer:1982}, was
discovered in 1982, and during the next decade or so almost every
year a new one was reported. In the recent years the situation
changed considerably with the discovery of an anomalously large
population of millisecond pulsars in globular
clusters~\cite{Weber:1999a}, where the density of stars is roughly
1000 times that in the field of the galaxy and which are therefore
very favorable sites for formation of rapidly rotating neutron stars
which have been spun up by the means of mass accretion from a binary
companion. Presently more than 700 pulsar have been reported, and
the detection rate is rather high.

In 2006 Hessels et al.~\cite{Hessels:2006ze} reported the discovery
of a very rapid pulsar J1748-2446ad, rotating at $\nu=716Hz$ and
thus breaking the previous record (of $641Hz$). However, even this
high rotational frequency is too low to affect the structure of
neutron stars with masses above $1M_{\odot}$~\cite{Bejger:2006hn}.
Such pulsars belong to the slow-rotation regime since their
frequencies are considerably lower than the Kepler (mass-shedding)
frequency $\nu_k$. (The mass-shedding, or Kepler, frequency is the
highest possible frequency for a star before it starts to shed mass
at the equator.) Neutron stars with masses above $1M_{\odot}$ enter
the rapid-rotation regime if their rotational frequencies are higher
than $1000Hz$~\cite{Bejger:2006hn}. A recent report by Kaaret et
al.~\cite{Kaaret:2006gr} suggests that the X-ray transient XTE
J1739-285 contains the most rapid pulsar ever detected rotating at
$\nu=1122Hz$. This discovery has reawaken the interest in building
models of rapidly rotating neutron stars~\cite{Bejger:2006hn}.

\begin{figure}[t!]
\begin{center}
\includegraphics[width=5.cm,height=5.cm]{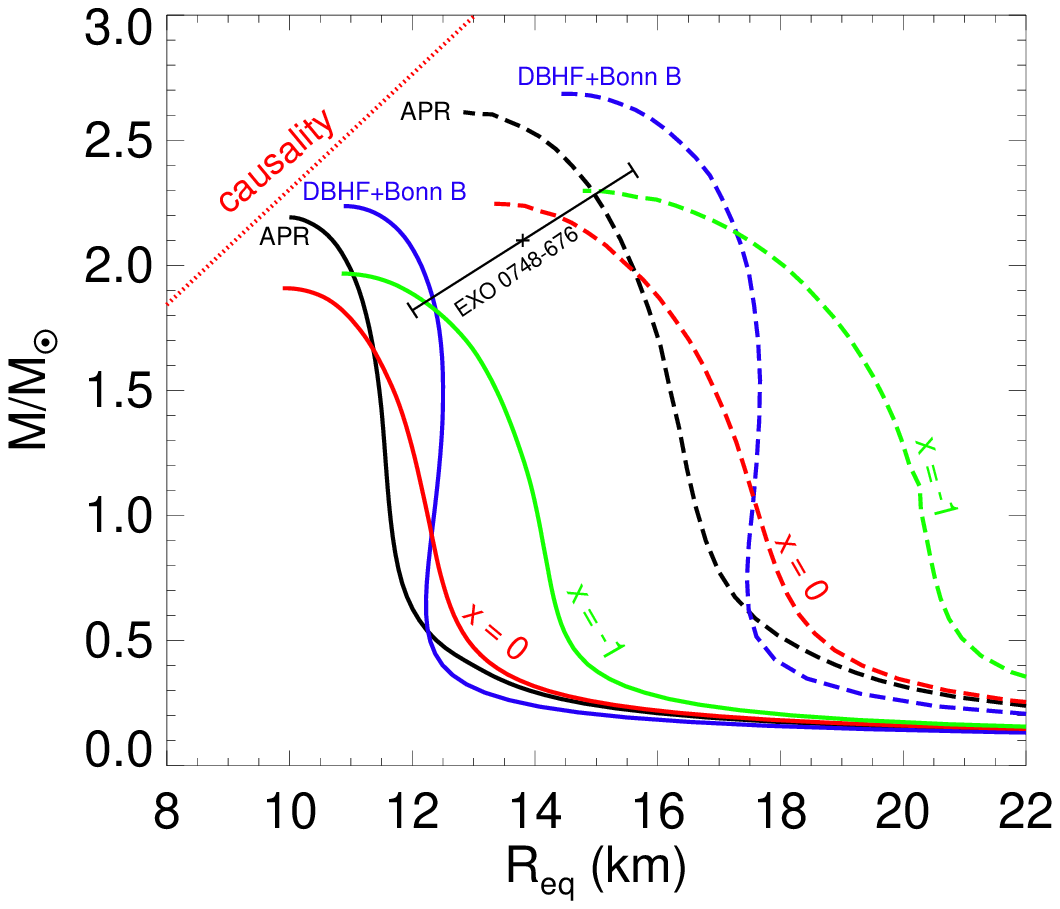}
\includegraphics[width=5.cm,height=5.cm]{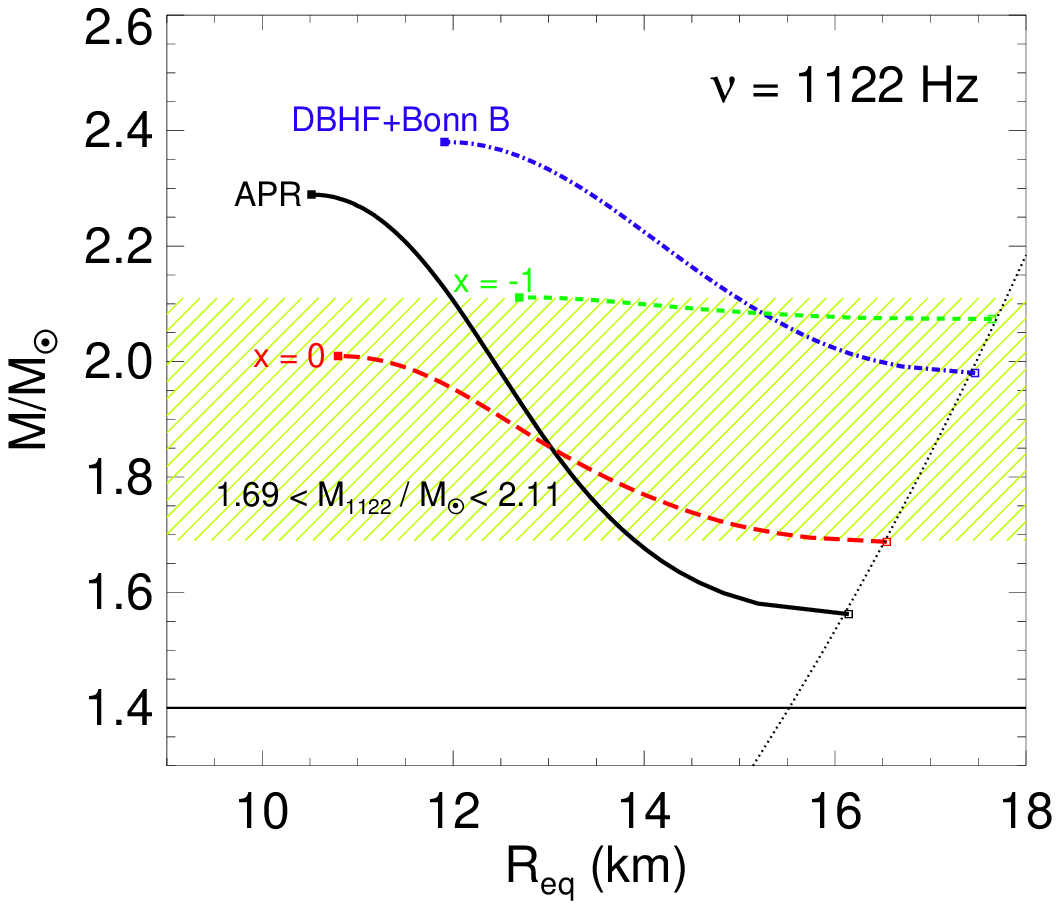}
\caption{{\protect\small Left panel: Mass-radius relation. Both
static (solid lines) and Keplerian (broken lines) sequences are
shown. The $1-\sigma$ error bar corresponds to the measurement of
the mass and radius of EXO 0748-676~\cite{Ozel:2006km}; Right panel:
Gravitational mass versus circumferential radius for neutron stars
rotating at $\nu=1122Hz$.}}
\end{center}
\label{mr}
\end{figure}

\begin{table}[!b]
\caption{Maximum-mass rapidly rotating models at the Kepler
frequency $\nu=\nu_k$.}
\begin{center}
\begin{tabular}{lcccc}
EOS &  $M_{max}(M_{\odot})$ & Increase (\%) & $\epsilon_c(\times 10^{15}g\hspace{1mm}cm^{-3})$ & $\nu_k(Hz)$\\
\hline\hline
MDI(x=0)     & 2.25    &  15  & 2.59 & 1742\\
APR          & 2.61    &  17  & 2.53 & 1963\\
MDI(x=-1)    & 2.30    &  14  & 2.21 & 1512\\
DBHF+Bonn B  & 2.69    &  17  & 2.06 & 1685\\
\hline
\end{tabular}
\end{center}
{\small The first column identifies the equation of state. The
remaining columns exhibit the following quantities for the maximally
rotating models with maximum gravitational mass: gravitational mass;
its percentage increase over the maximum gravitational mass of
static models; central mass energy density; maximum rotational
frequency.}
\end{table}

Applying several nucleonic equations of state (see previous section)
and the $RNS$\footnote{Thanks to Nikolaos Stergioulas the $RNS$ code
is available as a public domain program at
http://www.gravity.phys.uwm.edu/rns/} code developed and made
available to the public by Nikolaos
Stergioulas~\cite{Stergioulas:1994ea,Stergioulas:2003yp}, we
construct one-parameter 2-D stationary configurations of rapidly
rotating neutron stars (for details see Ref.~\cite{Krastev:2007wh}).
The computation solves the hydrostatic and Einstein field equations
for mass distributions rotating rigidly under the assumptions of
stationary and axial symmetry about the rotational axis, and
reflection symmetry about the equatorial plane.

The effect of ultra-fast rotation at the Kepler (mass-shedding)
frequency is examined in the left panel of Fig.~5 (see also Table 1)
where the stellar gravitational mass is given as a function of the
{\it equatorial} radius. Predictions are shown for both static
(non-rotating) and rapidly rotating stars. We observe that the total
gravitational mass supported by a given EOS is increased by rotation
up to 17\% (see Ref.~\cite{Krastev:2007wh}). At the same time, the
circumferential radius is increased by several kilometers while the
polar radius (not shown here) is decreased by several kilometers,
leading to an overall oblate shape of the rotating star.

Models of neutron stars rotating at $1122Hz$~\cite{Kaaret:2006gr}
are shown in Fig.~5 (right panel). Stability with respect to the
mass-shedding from equator implies that at a given gravitational
mass the equatorial radius $R_{eq}$ should be smaller than
$R_{eq}^{max}$ corresponding to the Keplerian
limit~\cite{Bejger:2006hn}. On the other hand, the stellar sequences
are terminate at $R_{eq}^{min}$ where the star becomes unstable
against axial-symmetric perturbations. In Fig.~5 (right panel) we
observe that the range of the allowed masses supported by a given
EOS for rapidly rotating neutron stars becomes narrower than the one
of static configurations. This effect becomes stronger with
increasing frequency and depends upon the EOS. Since predictions
from the $x=0$ and $x=-1$ EOSs represents the limits of the neutron
star models consistent with the nuclear data from terrestrial
laboratories, we conclude that the mass of the neutron star in XTE
J1739-285 is between 1.7 and $2.1M_{\odot}$.

\section{Summary}

\quad We have presented an overview of our recent studies of
effective interactions in dense neutron-rich matter and their
applications to problems with astrophysical significance. The DBHF
calculation of properties of spin-polarized neutron matter has been
extended to high densities. Although no transition to a
ferromagnetic phase is actually seen up to 10$\rho_0$, the observed
behavior of the spin-symmetry energy suggests that such transition
may be possible at much higher densities. Applying the
gravitochemical heating formalism developed by Jofre et
al.~\cite{Jofre:2006ug} and the EOS with constrained symmetry
energy, we have provided a limit on the possible time variation of
the gravitational constant $G$ in the range $(4.5-21)\times
10^{-12}yr^{-1}$. Our findings also allowed us to constrain the mass
of the neutron star in XTE J1739-285 to be between 1.7 and
$2.1M_{\odot}$.

In closing our discussion we would like to emphasize that further
progress of our understanding of properties of dense matter can be
achieved through coherent efforts of experiment, theory/modeling,
and astrophysical observations.

\section*{Acknowledgments}
We would like to thank Rodrigo Fern\'{a}ndez and Andreas Reisenegger
for helpful discussions and assistance with the numerics of the
gravitochemical heating method. We also thank Fiorella Burgio for
providing the hyperonic and hybrid EOSs and Wei-Zhou Jiang for
helpful discussions. The work of Plamen G. Krastev, Bao-An Li and
Aaron Worley was supported by the National Science Foundation under
Grant No. PHY0652548 and the Research Corporation under Award No.
7123. The work of Francesca Sammarruca was supported by the U.S.
Department of Energy under grant number DE-FG02-03ER41270.

\end{document}